# Automated Test Case Generation using Petri Nets

Jai Manral

NTT DATA FA Insurance Systems (NDFS), Bangalore, India

jai.manral@nttdata.com

*Abstract— Software testing is the process of determining the precision, quality, completeness and security of the software systems. An important step in testing software is the generation of test cases, whose quality plays a vital role in determining the time for testing and subsequently its cost. In this research, it is shown that both structural and behavioural diagrams can be used to represent specifications in a single model using High Level Petri Nets (HLPN). This research focuses on automated generation of test models from Petri nets. Moreover, generating consistent formal models (HLPN) from informal models (UML) is the highlight of this research.*

## I. INTRODUCTION

Unified Modeling Language (UML) is a standard modelling language used to describe the analysis and design specification of software [1]. UML is mostly used in object oriented system and is designed to incorporate current best practices in the field of software engineering. UML can be used with almost all software development methodologies or life cycle and thus provide an easy way to model a human-readable representation of a system in a visual format, that can easily be rendered by computers. The Unified Process (UP) is a methodology that uses UML as the underline visual modeling syntax and defines the workers, activities and artifacts needed to model the system [2]. The latest version of UML allows modeling the behaviour and structure of the system, so the complete system can be represented with one or more UML diagrams. Some of the examples of UML structured diagrams are class diagram, deployment diagram, component structure diagram, etc. Whereas the examples for behaviour diagrams are interaction diagram, state transition diagram, use case diagram and activity diagram. For the purpose of this research class diagram and sequence diagram will be used to generate the Petri Net representation of the system which in turn will be used to extract the test cases. The other diagrams used were Use Case Diagram and Activity Diagram.

### A. Class Diagram

The class diagram represents the objects of the system, its attributes and operations. The diagram shown in below Figure 1 is a structural representation of objects of the proposed system and their relationships. As classes are the units of objects oriented programming this diagram can be used to create the structure of the proposed system.

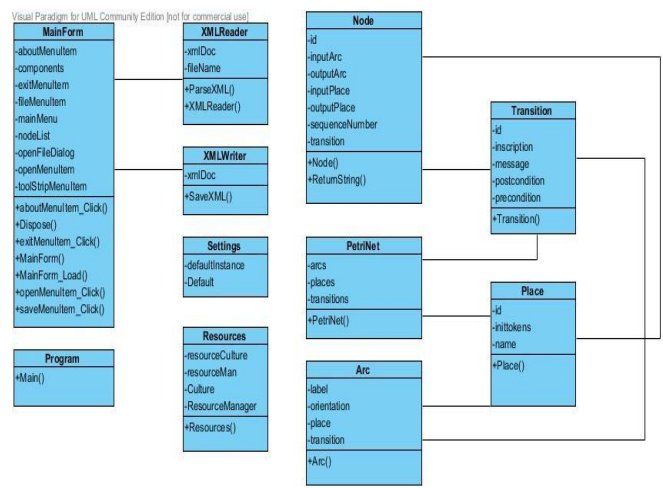

Fig 1: Class Diagram ATCGPNET

### B. Use Case Diagram

The use case diagram shown below Figure represents the interaction of user with the system and the overall processes of the system. This can be used to specify the processes and user/system interactions.

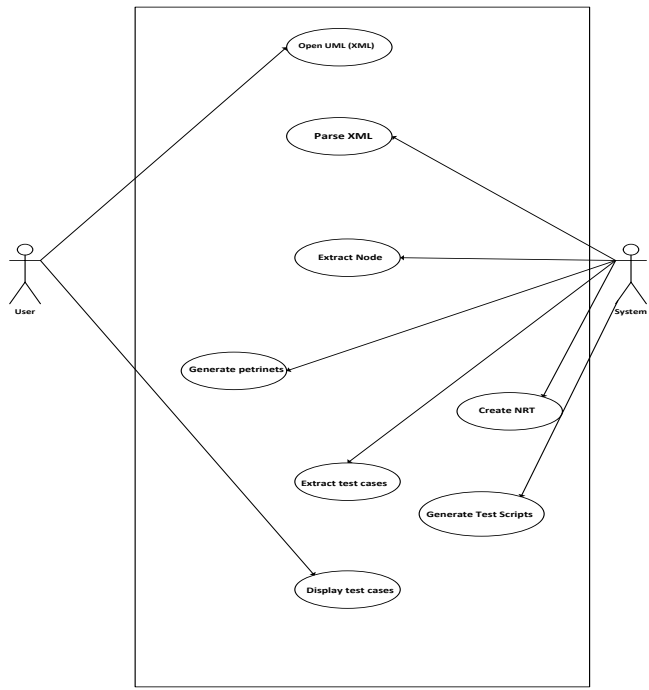

Fig 2: Use Case Diagram ATCGPNET

## C. Activity Diagram

Activity diagram shown in the below Figure represents the work flow of stepwise activities of the system. This diagram is used to describe the business and operation work flow in a step by step manner.

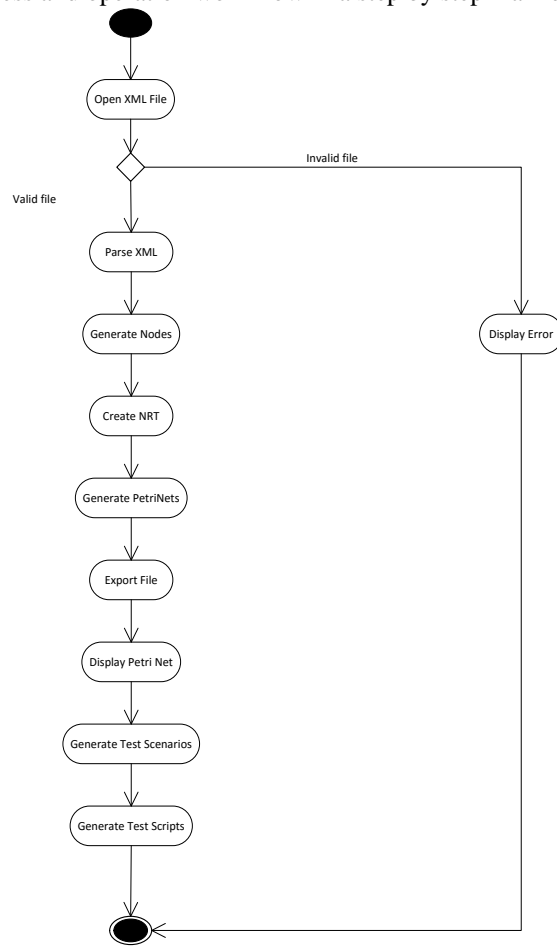

Fig 3: Activity Diagram ATCGPNET

*D. Sequence Diagram*

The sequence diagram shown in the Figure 4 represents the sequence of events or message passed between the classes of the system. This diagram is essential at representing the interaction/behaviour of the system when different processes are fired. Multiple Sequence diagrams are used to represent complex system where the system is designed to be used for a number of different functions like software systems.

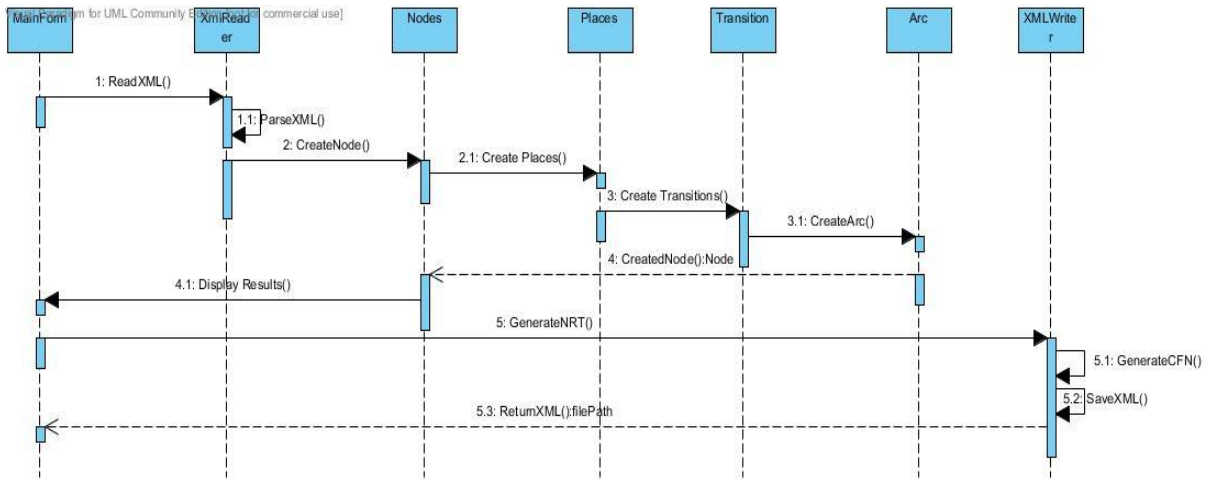

Fig 4: Sequence Diagram ATCGPNET

## II. IMPLEMENTATION

This section explains the various tools and techniques used for implementing the current research. The proposed solution named as ATCGPNET (Automated Test Case Generation using Petri Net) is used to extract the specification from XML file considering the software structural and behavioural specifications. The process used includes, extracting the specifications from UML and creating the nodes of the Petri Net. Then a NRT (Node Relationship Table) is created that represents the relationship of each node with all the other nodes of the system. Finally the Petri Net is generated by combining the Nodes together using the CFN (Combined Fragment Net) and NRT. The last step, generation of test cases from Petri Net is done using MISTA (Model-based Integration and System Testing Automation). The software specifications are created using a tool called Visual Paradigm. Details about Visual Paradigm are given in the sub section A below. Details about MISTA (Model-based Integration and System Testing Automation) and its usage in the current work are given in sub section B. Lastly, the details about the developed system are given under section C.

### A. Visual Paradigm for UML 10

Visual Paradigm is a tool that supports the modeling of systems using UML 2.0, SysML, BPMN and XMI. This tool was chosen due it its ability to allow the end user (designer) to enter the specifications in a graphical manner (UML Diagrams) and format it in XML format, so that any software system can use the specifications to automate processes. The XML generated by Visual Paradigm contains in-depth detail about the requirements making this tool the best option for implementing the current work. Furthermore, the UML Diagrams, along with OCL can be exported to XML in a single file containing all the specifications in one place. This comes in handy as only one file needs to be imported to the system making it easy for the user using the system. Visual Paradigm provides a complete toolset which can be used by software designers and developers to generate the software specification models for the system under development. It can be used for requirements capturing, software planning, test planning and class and data modeling.

For the current study, Visual Paradigm was used to design the UML Model in a visual manner and then convert all the specification to XML. The specifications used were extracted from Class and Sequence Diagram, to make sure that both the system structural and behavioural specifications are captured, so that the complete system can be represented by the Model. The diagrams were created using the drag drop user interface provided by Visual Paradigm. The specification were converted to XML using the built in functionality of Visual Paradigm that exports all the specifications including the constraints and specifications extracted from UML Diagrams. The screenshots given below show the process of creating Class Diagrams and Sequence Diagrams in Visual Paradigm. The process of adding the OCL constraints are also highlighted in the screenshots below.

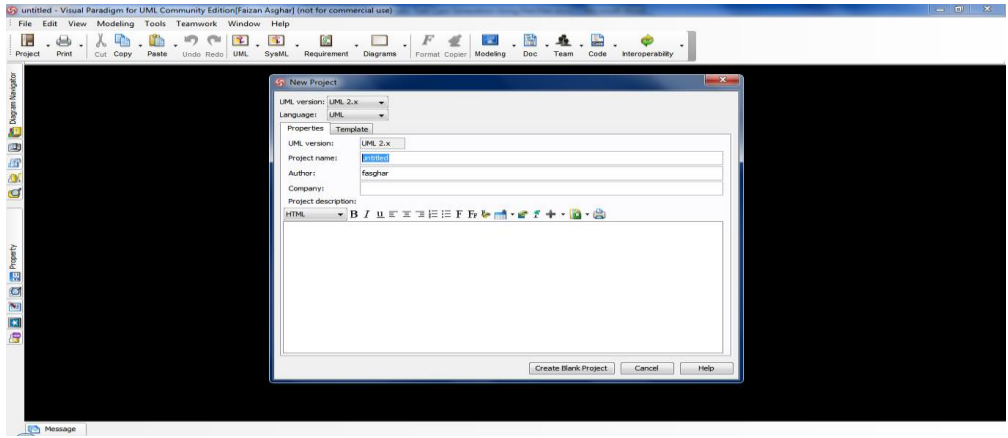

Fig 5: New Project screen

The figure above shows the process of creating a new project. The end user can specify the project and author details to create a new project. All project level information is specified at this step.

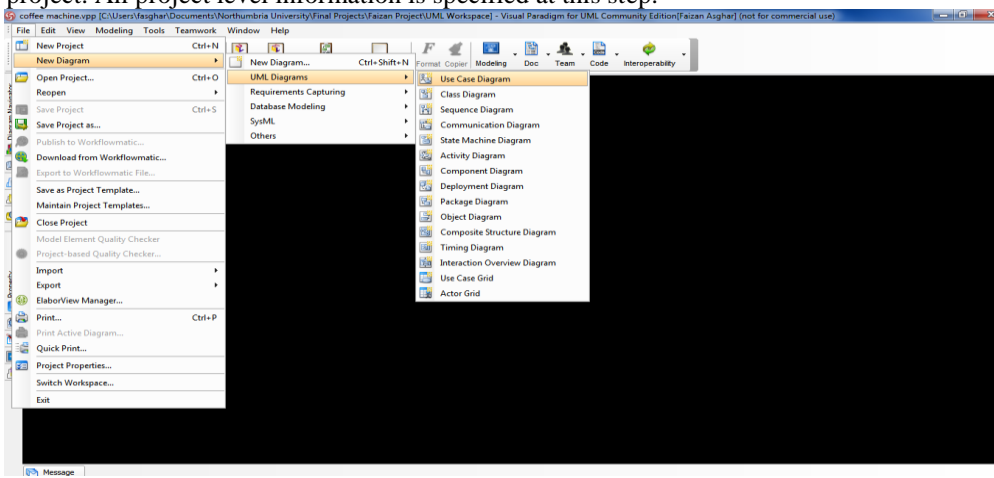

Fig 6: New UML Diagram screen

The screenshot above shows the process of creating a new UML Diagram. The end user can choose to select the relevant diagram that needs to be created at this step.

The next screenshot shows how to create a sequence diagram and explains of how to use different components of the software.

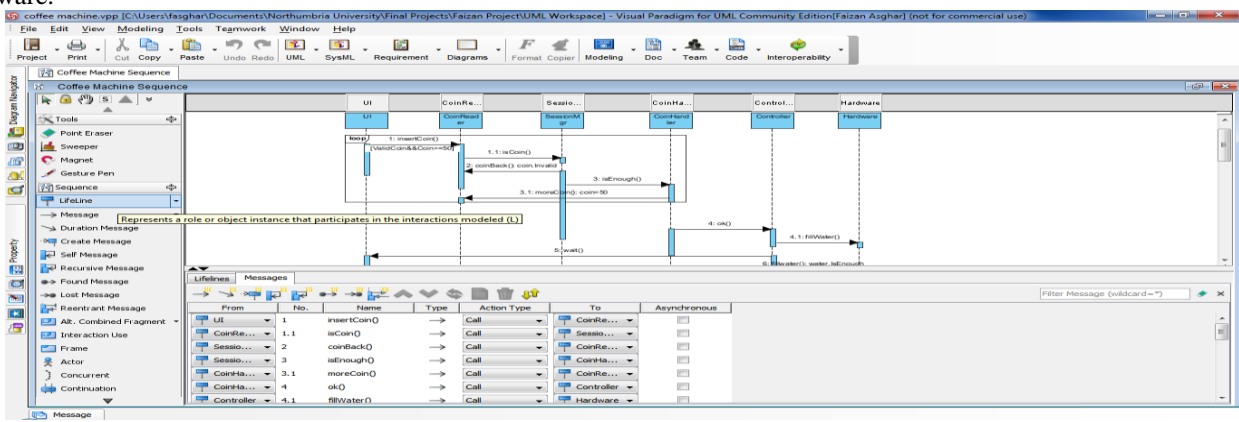

Fig 7: Create Sequence Diagram

The screenshot above shows all the different components that can be used to create a sequence diagram. These components include Message, Combined Fragments, Lifeline, Actors and many other components. The end user can select the component they want to add and then draw it on the screen at the desired location.

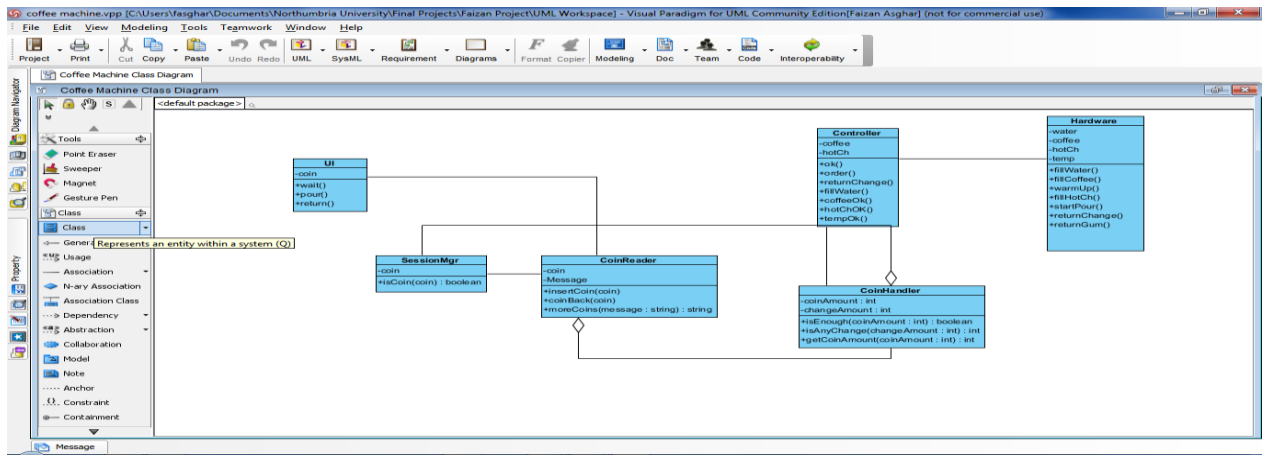

Fig 8: Create Class Diagram

The screenshot above represent a class diagram in Visual Paradigm. The end user can use the components on the toolbar on the left to create the desired class diagram structure. The components include Class, Generalization, Association, Model and many other to help represent the system in an efficient and orderly manner.

The screenshot below shows the process of adding OCL style specifications in the class diagram. The method pre and post conditions can be specified within the class diagram and the resulting XML file contains this info and enrich UML diagrams with required information that has no way of being represented in a visual manner. The end user can specify the constraints of the system in this fashion for all the constraints in the system.

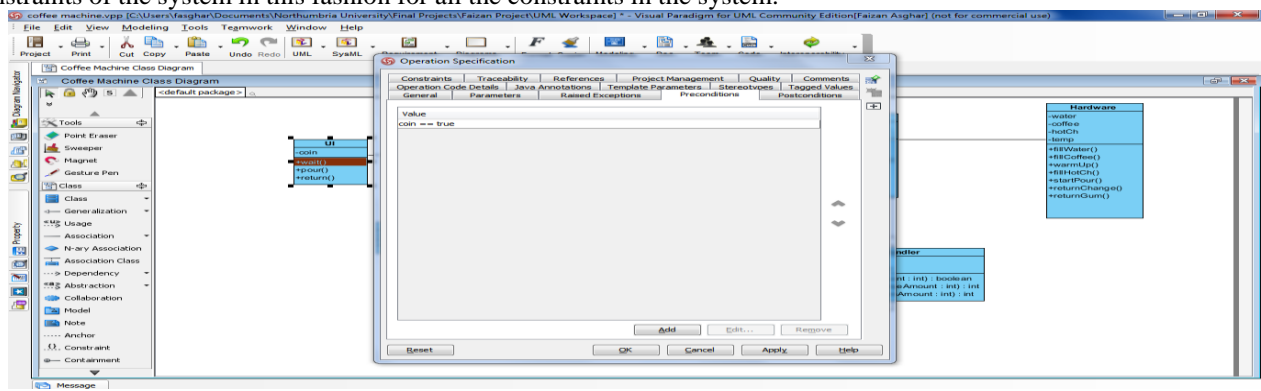

Fig 9: Add Constraints

### B. MISTA (Model-based Integration and System Testing Automation)

MISTA is a tool for model based test generation. It can generate test cases by extracting the specifications to generate a formal representation of the system using Petri Nets. It can also be used to validate the model and execute the test cases to generate the reachability graph. Furthermore, it creates the test scripts used to test the functionality in a variety of different languages and frameworks like C#, Java, NUnit, JUnit, etc. MISTA supports variety of functionality that can be used to test, modify and generate cases for the model.

For the purpose of current work, MISTA is used to visualize the Petri Nets created using the proposed approach and generate test cases along with the testing model. The Petri Net can be visualized, validated for syntax, simulated and run to make sure that the model is a complete representation of the software specification. The files generated using ATCGPNET is used as an input by MISTA for visualizing the software specification model using Petri Net. Furthermore, the test reachability graph and the model test cases can then be extracted from Petri Net by MISTA. Finally the Test cases generated can be converted into test scripts used to run in most of the popular frameworks like JUnit, NUnit and Selenium IDE. Most of the popular programming languages are also supported by MISTA like C#, Java, etc. The process of visualizing a Petri Net and extracting the test cases using MISTA is illustrated in the screenshots below.

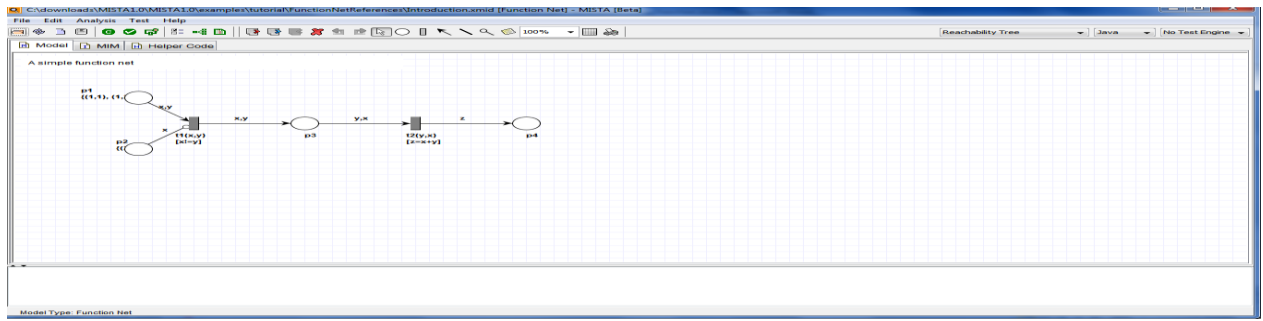

Fig 10: Open file MISTA

The screenshot above shows the sample Petri Net loaded from file using the File<-Open menu item. Once the file is loaded the Petri Net is displayed on the screen. The user can do a number of actions on the Petri Nets using the options in the toolbar on top. Some of the actions are displayed below:

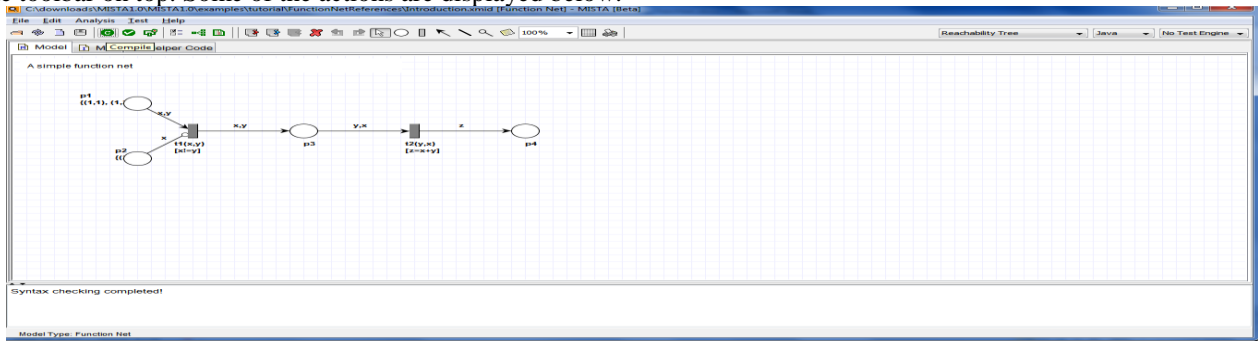

Fig 11: Compile Petri Net

The Compile toolbar button shown above is used to check the Petri Net for syntax errors. The results are shown in the console located at the bottom on the screenshot above.

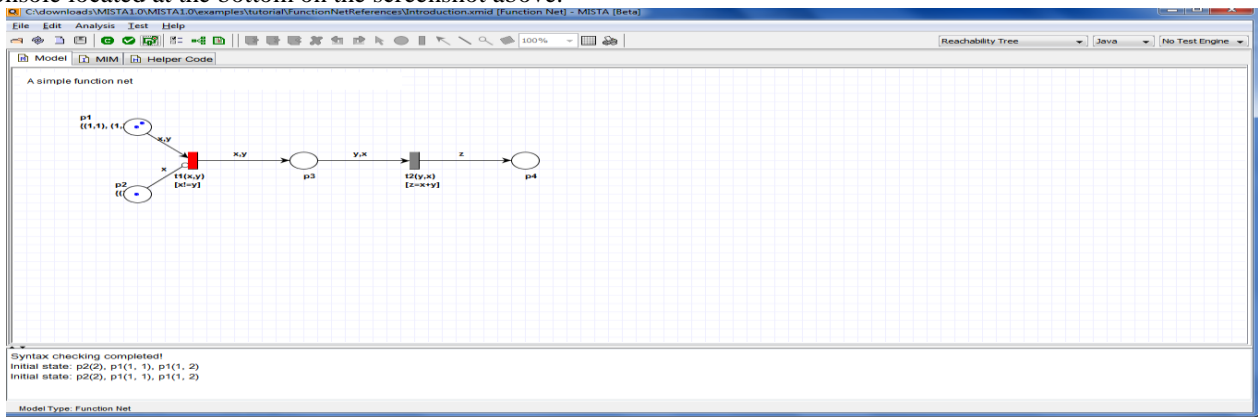

Fig 12: Simulate Model

When the Simulate toolbar button is pressed, net simulation is started and tokens are fired from the initial place. The screenshot given below shows the control panel for the simulation. The user can select the events, parameters and fire events to simulate different paths of the Petri net.

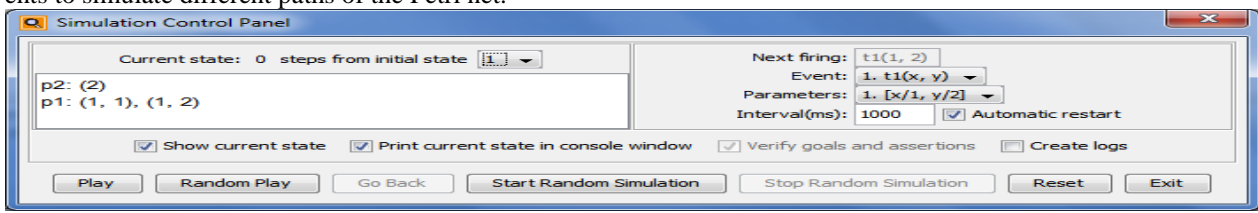

Fig 13: Simulate Control Panel

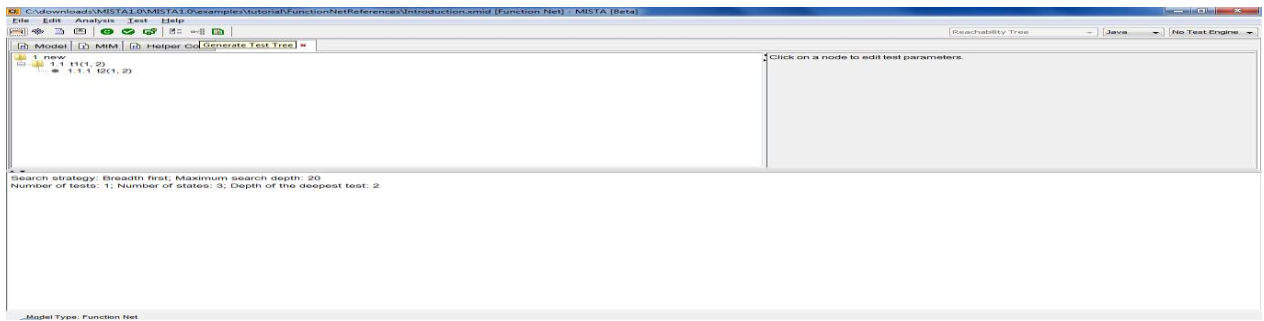

Fig 14: Test Tree

The figure shown above shows the testing tree generated from the model by pressing the Generate Test Tree button on the toolbar.

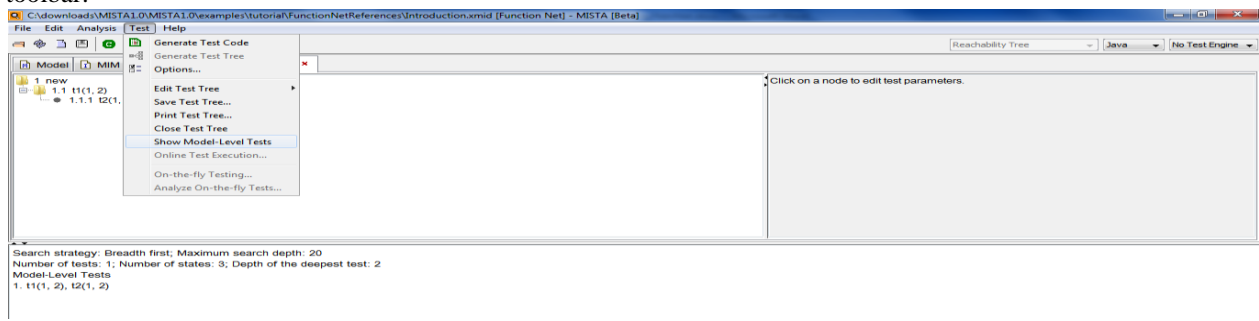

Fig 15: Model-Level Test Cases

Once the Test Tree is generated the Model Test cases can be extracted from the tree. This can be done by selecting the Show Model-Level Tests menu item under the Test menu. The results are shown in the console at the bottom of the window.

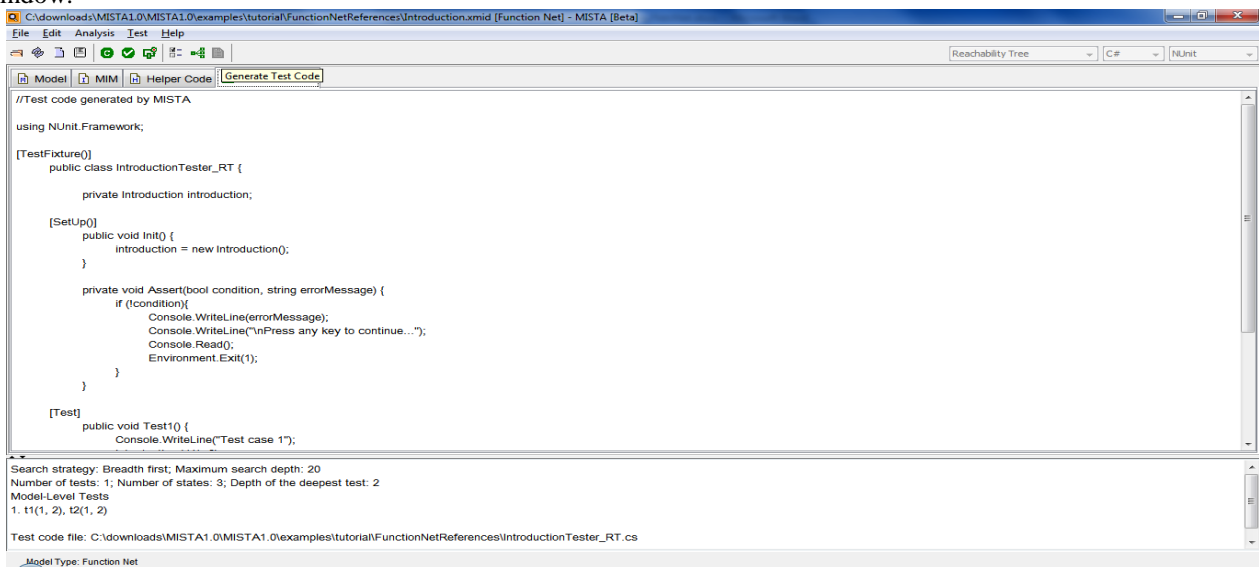

Fig 16:Test Code

The Test Code can be generated using the Generate Test Code option on the toolbar. This generates a code file containing all the code required to test the captured test cases in the programming language of choice. The language and the testing frameworks can be changes from the top right of the screen using the appropriate dropdown. The screenshot above shows code generated for NUnit and C#. The file path is also shown in the console.

### C. ATCGPNET (Automatic Test Case Generation using Petri Nets)

The proposed solution ATCGPNET (Automatic Test Case Generation using Petri net) is a tools used for extracting the software structural and behavioral specification from UML Class Diagram, Sequence Diagram with the help of OCL (Object Constraint Language). The system accepts an XML file containing the UML Diagrams and OCL Specification in XML format. The XML specification file is then parsed and the required specifications are extracted. These specifications are then used to create the places, transitions, arcs and tokens for Petri Net model. Nodes are created from the specifications. Each Node has an input place, output place, input arc, output arc, transition and a list of tokens at each place. Once node creation process is complete, a NRT (Node Relationship Table) is constructed using the associations specified in class diagram and the message sequence specified by the sequence diagram. Furthermore, Combined Fragments are linked together using the NRT to generate a Combined Fragment Tree which is in turn combined with the nodes outside combined fragments in the sequence diagram to create a Petri Net. Finally the data is exported in XML format to be used in MISTA for visualization and test case generation purposes.

The tool was developed in C# using .NET framework 4.0. Visual Studio 2010 Ultimate was used to develop the software system. LINQ to SQL was used to query the XML files to extract the specification from UML and export the Petri Net model to XML. LINQ to SQL provides a convenient way for developers to query specific objects from XML and makes the process of information retrieval to and from XML easy. Various Classes and design patterns were used to make the design of the system flexible and to allow for change. Appendix 2 contains all the code developed for the current study. The screenshots below show the process of loading the XML, creating the net and exporting it to format that can be opened by MISTA for further processing.

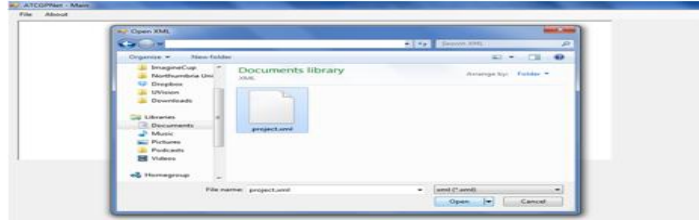

Fig 17: Open Specification

The screenshot above shows the file open screen of ATCGPNET, the user can select the file generated using Visual Paradigm and the Nodes will be extracted from the specifications. Once done the nodes are displayed on the screen as shown in the screenshot below:

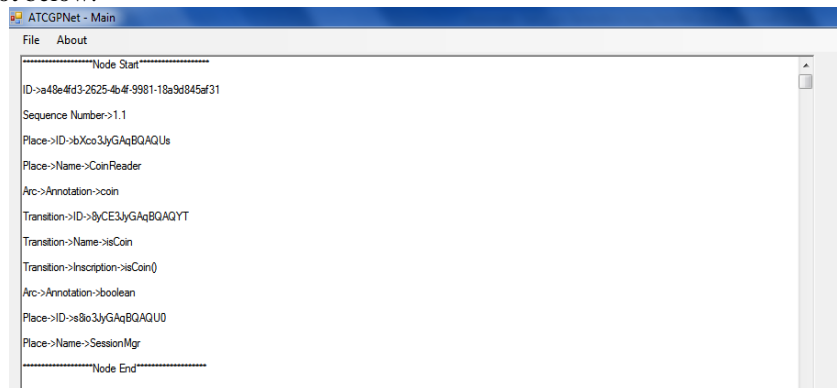

Fig 18: Show Nodes

The user can add or modify any specifications extracted from XML if required otherwise saving the file will generate the petri net model and save a new file in the same directory. The screenshot of the generated file is giving below:

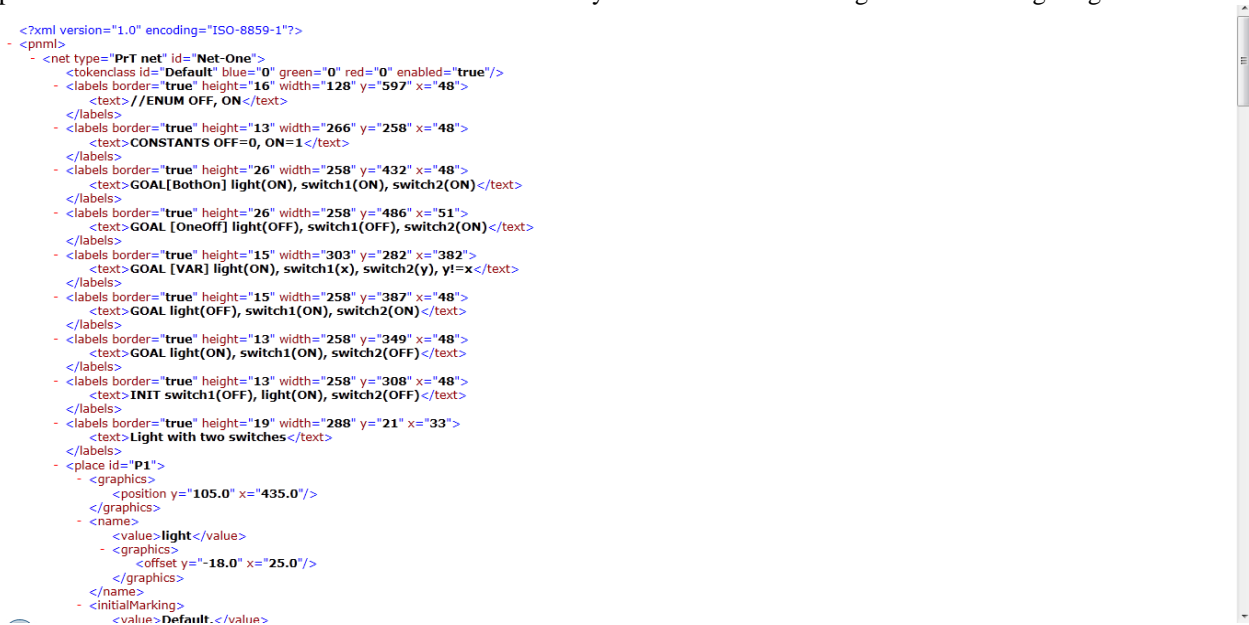

Fig 19: Resulting XML file

III. RESULTS

This chapter uses an example to describe the working on the proposed system. The example used is for a login process. The main objective of this chapter is to explain the working of the system, so a simple example is used to allow the reader to easily understand the internal working of the system. Subsection A presents the results generated by using this example, and subsection B analyses the results. Finally subsection C provides the summary for this chapter.

*A. Results*

This section displays the results of the proposed implementation method. The example of a simple login system is used to describe the working on the system. The figure below shows a class diagram for the login system. A user enters username and password to log in to the system. The system validates the user and if the user is valid allows the user to login.

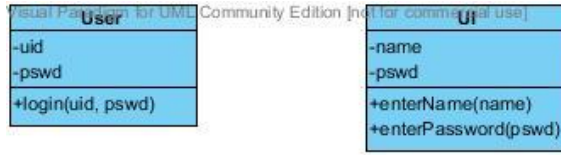

Fig 20: Login Class Diagram

The diagram below is the sequence diagram for the login system; this shows the flow of events for the login process. These two diagrams along with OCL is exported to XML and in turn is used by ATCGPNET to generate a petri net model.

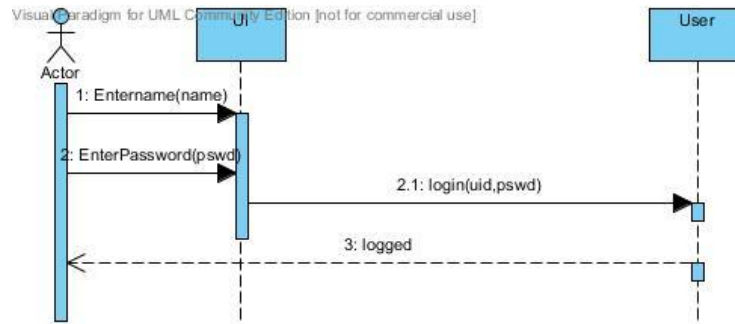

Fig 21:Login Sequence Diagram

The screenshot of the xml file given below is generated from ATCGPNET tool; the specifications were parsed by the tool and then converted into nodes. Finally the nodes were linked together to create the Petri Net. Once the net model was created it was then again converted to XML using PNML and opened using MISTA.

```xml
<?xml version="1.0" encoding="ISO-8859-1"?>
<pnml>
  <net type="PrT net" id="Net-One">
    <tokenclass id="Default" blue="0" green="0" red="0" enabled="true"/>
    <labels border="true" height="13" width="539" y="316" x="83">
      <text>INIT name(UID), password(PSWD)</text>
    </labels>
    <place id="P2">
      <graphics>
        <position y="105.0" x="225.0"/>
      </graphics>
      <name>
        <value>P2</value>
        <graphics>
          <offset y="35.0" x="-5.0"/>
        </graphics>
      </name>
      <initialMarking>
        <value>Default,</value>
        <graphics>
          <offset y="0.0" x="0.0"/>
        </graphics>
      </initialMarking>
      <capacity>
        <value>0</value>
      </capacity>
    </place>
    <place id="P3">
      <graphics>
        <position y="105.0" x="420.0"/>
      </graphics>
      <name>
        <value>P3</value>
        <graphics>
          <offset y="35.0" x="-5.0"/>
        </graphics>
      </name>
      <initialMarking>
        <value>Default,</value>
        <graphics>
          <offset y="0.0" x="0.0"/>
        </graphics>
      </initialMarking>
```

Fig 22: XML Generated From ATCGPNET

Once the file is opened you can see the model visualization as shown in the screenshot below, the simulation can be started at this point to visualize the working of the net.

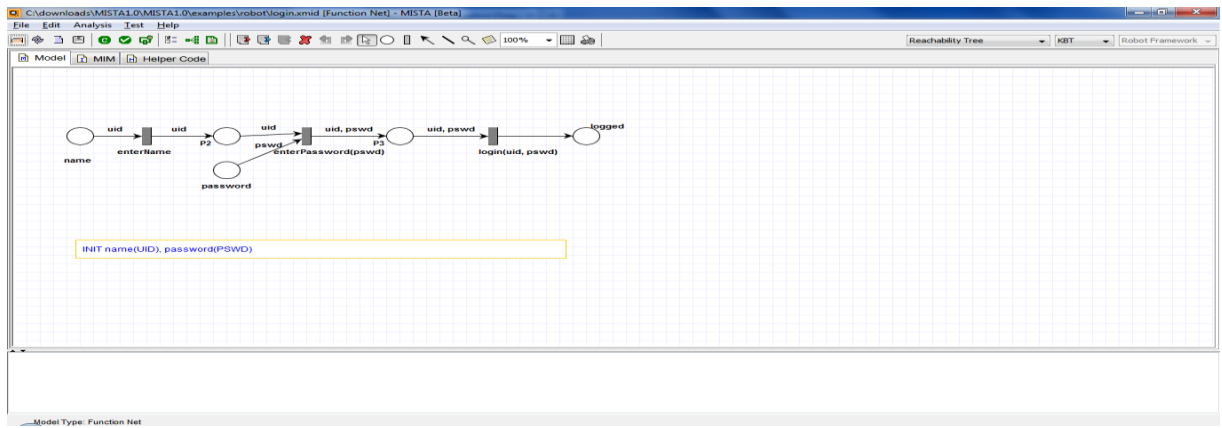

Fig 23: Petri Net from XML

The Test Tree is a tree structure of all the test scenarios, it can be generated for the login process above as shown in the screenshot below:

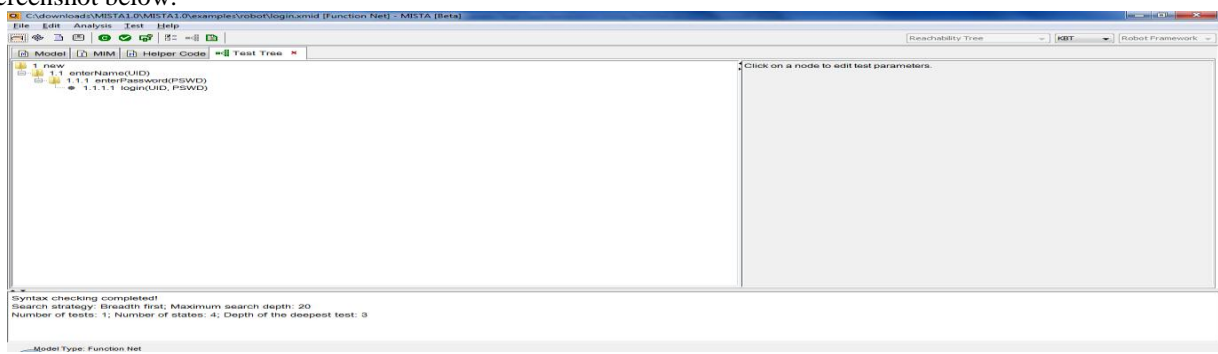

Fig 24: Test Tree

The Test case scenarios are shown below and have been generated from MISTA using the model created using ATCGPNET.

**Model-Level Tests**
1. enterName(UID), enterPassword(PSWD), login(UID, PSWD)

Fig 25: Test Scenarios

The screenshot below shows the test scripts that have been generated for the login process. These test scripts were generated for NUnit and C#. There are a number of choices available for programming languages and testing frameworks, which can be used to generate the test script that further facilitates the user by not restricting the choice of programming language.

```
//Test code generated by MISTA

Library        Selenium Library
Resource       LoginKeywords.txt
Force Tags     FunctionalTest
Default Tags   ValidTest
using NUnit.Framework;

[TestFixture()]
    public class loginTester_RT {

        private login login;

        [SetUp()]
            public void Init() {
                login = new login();
            }

            private void Assert(bool condition, string errorMessage) {
                if (!condition) {
                    Console.WriteLine(errorMessage);
                    Console.WriteLine("\nPress any key to continue...");
                    Console.Read();
                    Environment.Exit(1);
                }
            }
```

Model-Level Tests
1. enterName(UID), enterPassword(PSWD), login(UID, PSWD)

Test code file: C:\downloads\MISTA1.0\MISTA1.0\examples\robot\loginTester_RT.cs

Fig 26: NUnit Test Script in C#

*B. Analysis*

The example is a simple illustration of the working system and is displayed here to make the working of the system earlier to understand. The test scenarios generated using complex system during the research showed the same test quality and correctness as [3], but test scripts were available, which are a great help for automating the testing process.

The results found for the coffee machine sequence diagram used in [3] are displayed below, which are identical to the ones generated by latter.

| Senarios | States |
|----------|--------|
| Sc1 | m0 ->T1 -> m1 |
| Sc2 | m0 ->T1 -> m1->T2->m0 |
| Sc3 | m0->T3->m2 |
| Sc4 | m0->T3->m2->T4->m3 |
| Sc5 | m0->T3->m2->T4->m3->T5->m4 |
| Sc6 | m0->T3->m2->T4->m3->T5->m4->T6->m7 |
| Sc7 | m0->T3->m2->T4->m3->T5->m4->T7->m8 |
| Sc8 | m0->T3->m2->T4->m3->T6->m5 |
| Sc9 | m0->T3->m2->T4->m3->T7->m6 |
| Sc10 | m0->T3->m2->T4->m3->T7->m6->T5->m8 |
| Sc11 | m0->T3->m2->T4->m3->T6->m5->T5->m7 |

Fig 27: Testing Scenarios

*C. Summary*

This chapter uses the Login process example to show the working of the system and display the results generated. First a brief explanation is given to show how the login process works, then class and sequence diagrams are displayed which are generated using Visual Paradigm. Then, the Petri Net model XML document is displayed that has been generated using the proposed solutions. Finally the steps required for visualization of Petri Net is shown. At the end, the results gathered are analysed to make sure the result were as expected.

## IV. CONCLUSIONS

The main objective of the research was to extend the existing research done by [3] in such a way to add quality and improve the way the test cases are presented to the user, so that the generated test cases can add more value to the end user. Test case generation is one of the most important stages of software testing. Majority of project time is spent on manually creating the test cases by following the system specification model created by software designers. Software testing can be carried out at two different points in the Software Development Life Cycle (SDLC); the first once is after the completion of development phase and secondly after the model design phase. Software defects also known as bugs are difficult to fix and require more time ending up to be costly, if they are caught at a later phase in the SDLC. This is due to the fact that current software are complex as they try to solve real world problems this makes it hard to figure out which part in the system has malfunctioned. To counter this fact a lot of techniques like design patterns are used by developers to write the code in such a fashion to make it flexible and easy to maintain, as software systems are prune to change. Test case generation at the model level provides a way to reduce the cost of software testing but the complexity of generating test cases is increased dramatically. The complexity can be reduced by generating the test cases that cover all/most of the software specification using a systematic approach.

The current research makes use of the latter approach to extract test cases from software specification model. The model needs to represent the structural and behavioral specification for it to represent all aspects of the software system. For this UML Diagrams are used. UML diagrams are divided into two main parts which are Structural and Interactional. Each UML diagrams represents a different view point of the system but Sequence Diagram and Class Diagram can be used to capture most of the structural and behavioral specifications of the system. So for the current study these two diagrams were used to extract the system specifications. The missing information in these diagrams which is required to generate consistent and high quality test cases was taken from OCL embedded within the Diagrams using Visual Paradigm. It is important to note here that UML is an informal way of representing the system and this can create inconsistency and informal nature of test cases that are generated using this approach. To eradicate this issue Petri Nets are used to represent the software specifications in a formal model. Once the Petri Net model is constructed, it can be used to validate the model, create test scenarios or test scripts from the resulting net. This allows in saving time, money and effort spent to generate test documents from software design specification. Furthermore, this allows the designers to test the model against the generated test cases to catch any faults early so that they can be rectified even before a single line of code is written.

## ACKNOWLEDGMENT

The author expresses heartfelt gratitude to the highly inspiring mentors from academia and industrial research, whose time, advice and guidance were extremely valuable and greatly appreciated, during the development of this research.